\documentstyle[11pt,newpasp,twoside,epsf]{article}
\def\edcomment#1{\iffalse\marginpar{\raggedright\sl#1\/}\else\relax\fi}
\reversemarginpar

\begin{document}
\title{Gigahertz-Peaked Spectrum Radio Sources in Nearby Galaxies}
 \author{Neil M. Nagar}
\affil{Arcetri Observatory, Largo E. Fermi 5, 50125 Firenze, Italy}
\author{Andrew S. Wilson}
\affil{University of Maryland, College Park, MD 20742, USA}
\author{Heino Falcke}
\affil{Max-Planck-Institut f\"{u}r Radioastronomie, Auf dem H\"{u}gel 69,
                53121 Bonn, Germany}
\author{James S. Ulvestad}
\affil{National Radio Astronomy Observatory, P.O. Box 0,
                Socorro, NM 87801, USA}
\author{Carole G. Mundell}
\affil{Liverpool John Moores University, Twelve Quays House, Egerton Wharf, 
       Birkenhead, Merseyside CH41 1LD, UK}

\begin{abstract}
There is now strong evidence that many LLAGNs contain accreting
massive black holes and that the nuclear radio emission is dominated
by parsec-scale jets launched by these black holes. 
Here, we present preliminary results on the 1.4~GHz to 667~GHz spectral 
shape of a well-defined sample of 16 LLAGNs. 
The LLAGNs have a falling spectrum at high GHz frequencies.
Several also show a low-frequency turnover with a peak in the 1-20~GHz 
range.
The results provide further support for jet dominance of
the core radio emission.
The LLAGNs show intriguing similarities with gigahertz-peaked
spectrum (GPS) sources.
\end{abstract}

\section{Introduction}

Low luminosity active galactic nuclei (LLAGNs), operationally defined as 
AGNs with nuclear H$\alpha$ luminosity 
%(L$_{H\alpha}$) 
$<$~10$^{40}$ erg~s$^{-1}$,
make up almost 50\% of all nearby bright galaxies
(e.g., Ho, Filippenko, \& Sargent 1997a). They are spectroscopically 
sub-classified  into Low Ionization Nuclear Emission Region nuclei (LINERs),
low luminosity Seyferts, and ``transition nuclei'', whose spectra are 
intermediate between Seyfert/LINER and HII region spectra.          

Evidence has been accumulating that some fraction of LLAGNs share 
characteristics in common with more powerful AGNs.
These similarities include the presence of nuclear compact radio cores
(Heckman 1980), water vapor megamasers (Braatz et al. 1997), 
nuclear point-like UV sources (Maoz et al. 1995; Barth et al. 1998),
broad H$\alpha$ lines (Ho et al. 1997b), and broader
H$\alpha$ lines in polarized emission than in total emission 
(Barth et al. 1999).
If LLAGNs are truly scaled down AGNs then the challenge is to explain their
much lower accretion luminosities. This requires either
very low accretion rates 
($\sim$10$^{-8}\,$L$_{\rm Edd}$; e.g. Falcke \& Biermann 1999)
or radiative efficiencies (the ratio of radiated energy to accreted mass)
much lower than the typical value of $\sim$10\%
(e.g.~Chapter~7.8 of Frank, King, \& Raine 1995) assumed for powerful AGNs.

Our radio surveys of the 96 nearest LLAGNs from the Palomar spectroscopic
survey  (Ho et al. 1997a) have found compact (150~mas) flat-spectrum radio 
cores in almost half of all LINERs and low-luminosity Seyferts 
(Nagar et al. 2000; 2002).
Follow-up observations detected parsec-scale radio cores in
all (16) LLAGNs with S$^{\rm VLA}_{\rm 2cm}$~$\geq$ 2.7~mJy, 
implying brightness temperatures 
$\ga$ 10$^8$~K (Falcke et al. 2000; Nagar et al. 2002).
The five nuclei with the highest core fluxes -
NGC~3031 (Bietenholz, Bartel, \& Rupen 2000),
NGC~4278 (Jones, Wrobel, \& Shaffer 1984; Falcke et al. 2000),
NGC~4486 (M~87; Junor \& Biretta 1995),
NGC~4374 (M~84; Wrobel, Walker, \& Bridle 1996; Nagar et al. 2002),
and NGC~4552 (M~89; Nagar et al. 2002) - all have pc-scale 
radio extensions, morphologically reminiscent of jets (e.g. Fig.~1).
The high brightness-temperatures rule out an origin for the radio
emission in star-formation processes. Thermal emission can also
be ruled out as it would imply two to four orders of magnitude higher
soft X-ray fluxes than seen in LLAGNs (see Falcke et al. 2000).

We are currently working on a high-resolution multifrequency radio
study of all 16 LLAGNs at D $<$ 19~Mpc with confirmed pc-scale 
radio cores. Results on the 5~GHz to 15~GHz radio spectra are 
published in Nagar, Wilson, \& Falcke (2001). 
Here, for the first time, we present preliminary results on the 
1.4~GHz to 667~GHz spectral shape.

\begin{figure}[t]
\plottwo{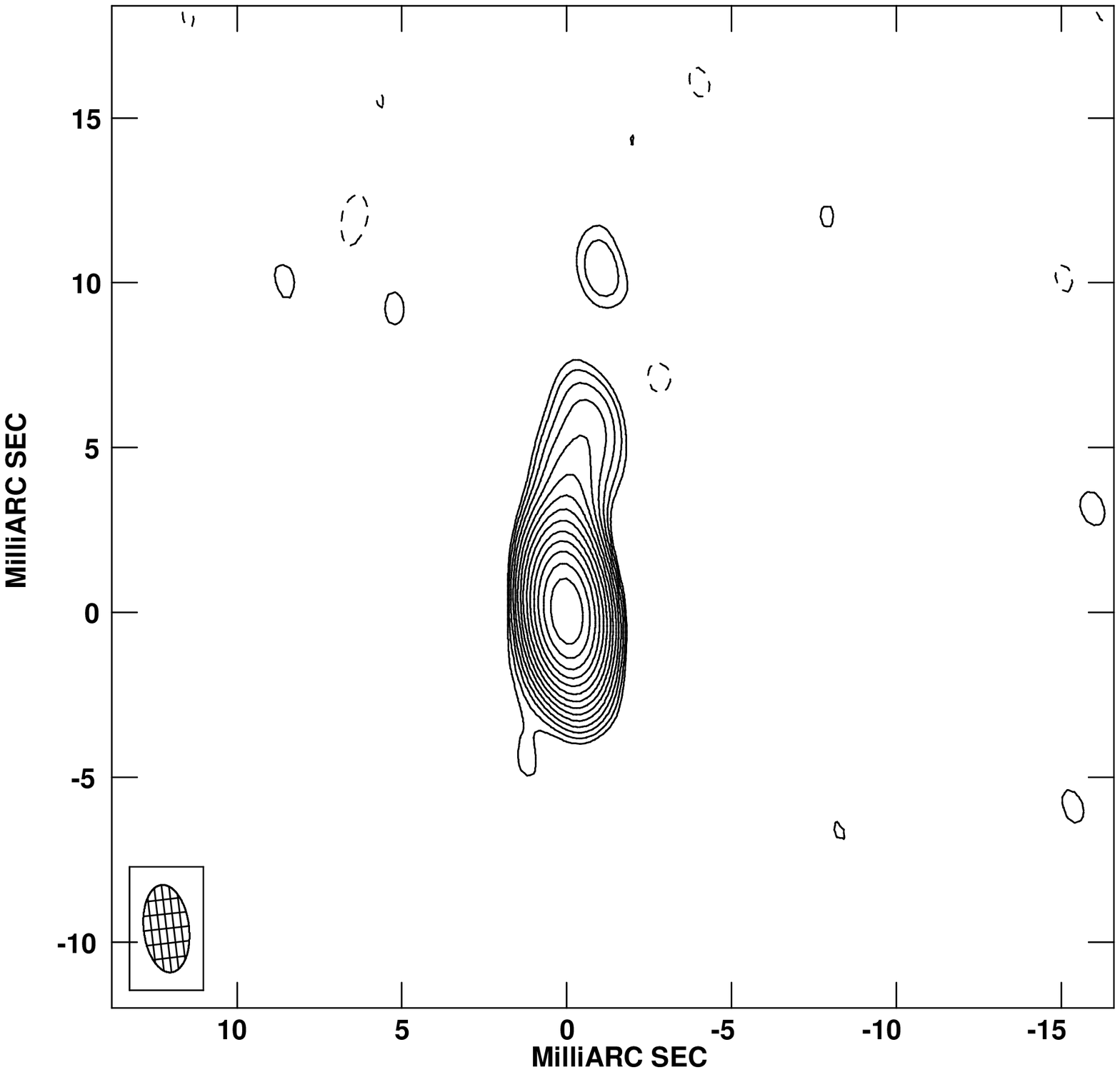}{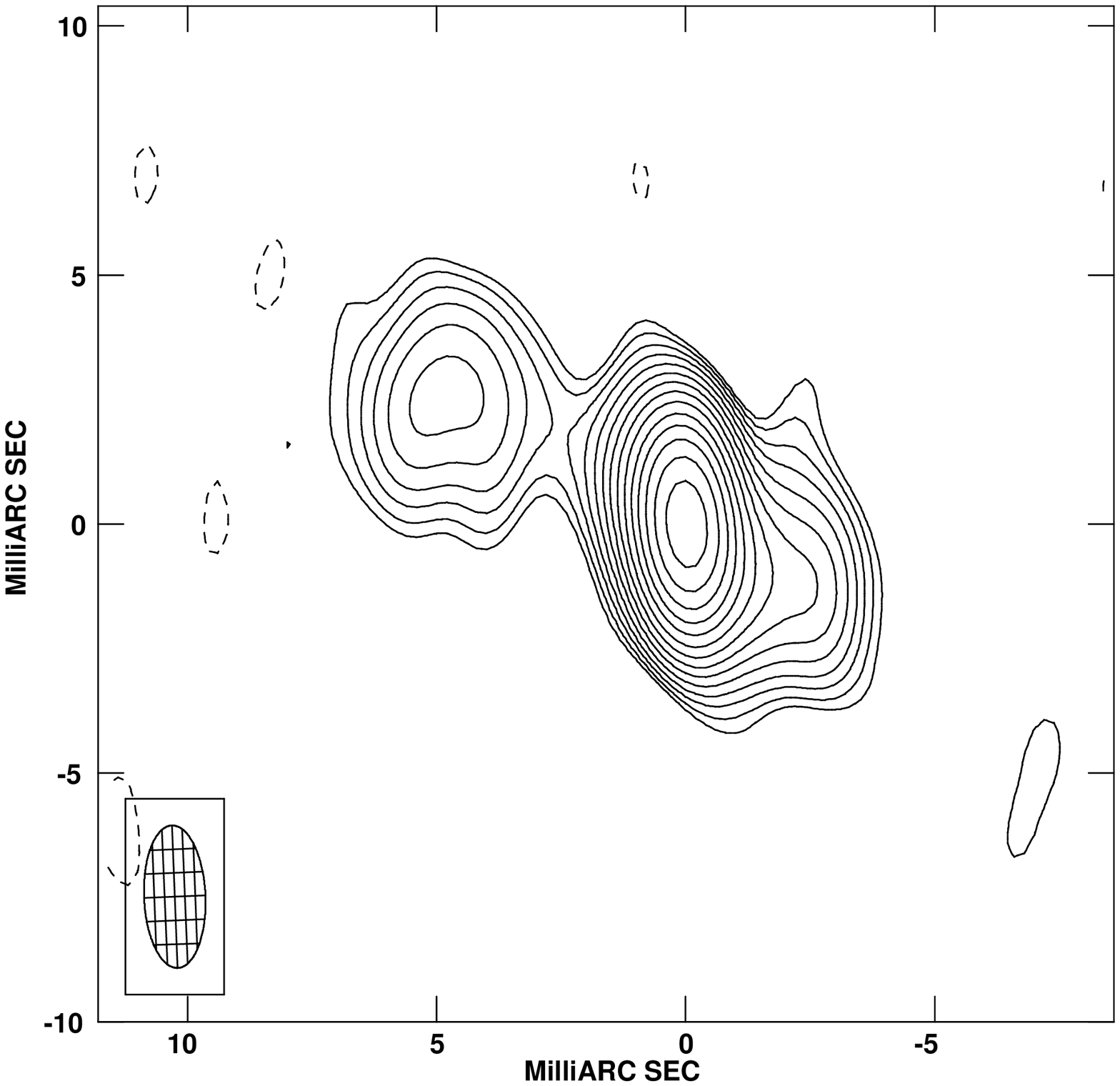}
\caption{5~GHz VLBA maps of NGC~4374 (left) and NGC~4552 (right).
 The contours are integer powers of $\sqrt{2}$, multiplied by the
 $\sim3\,\sigma$ noise level of 1.2~mJy for NGC~4374, and by the
 $\sim2\,\sigma$ noise level of 0.8~mJy for NGC~4552.
 The peak flux-densities are 152.6~mJy/beam and 93.8~mJy/beam,
 respectively. From Nagar et al. (2002).}
\end{figure}

\section{Observations} 
All sixteen nuclei have been observed simultaneously at 5~GHz, 
8.4~GHz, and 15~GHz with the VLA in September 1999, while the 
VLA was in A-array (results in Nagar et al. 2001). 
All nuclei were again observed at 8.4~GHz and 43~GHz in 
December 2000 with the VLA A-array, and the brightest eight nuclei 
were also observed at 22~GHz at this time. The repeat 8.4~GHz observations
were used to scale the flux measurements at 22~GHz and 43~GHz to the 
previous 5--15~GHz data; this, of course, assumes that the 5--43~GHz
spectral shape did not change between the two epochs.
We also observed eleven of the sixteen nuclei with MERLIN at 1.6~GHz
(0{\farcs}1 resolution) in December 1999. 
Thirteen of the 16 nuclei have been previously observed at
1.4~GHz (at 5{\arcsec} resolution) as part of the FIRST VLA radio 
survey (White et al. 1997) during various observing runs between 1997 and 1999.
Finally, we observed four of the sixteen nuclei at
353~GHz (850~$\mu$m) and 667~GHz (450~$\mu$m) with JCMT/SCUBA in
January 2001. Observations of $\sim$6 more nuclei with SCUBA are pending.

%The two epochs of VLA fluxes were tied together using the repeat 8.4~GHz
%observations. 
For the 5~GHz, 8.4~GHz, and 15~GHz
data, we were able to make matched 0{\farcs}5 resolution maps.
At 22~GHz and 43~GHz the much higher resolution (0{\farcs}08 
and 0{\farcs}05, respectively) and lower signal to noise, did not allow 
good maps to be made at 0{\farcs}5 resolution, and so we use the full 
resolution data in these cases. The radio cores are essentially point
sources at 22~GHz and 43~GHz, so it is unlikely that tapered maps will pick
up a significant amount of more extended emission.
%Given the smaller
%number of baselines at MERLIN, mapping is more difficult and hence
%these data have the highest uncertainty in the flux. 
At this stage our reduction of the MERLIN data are preliminary so we 
rely mainly on the 5{\arcsec} resolution, 1.4~GHz fluxes from the FIRST survey.
The SCUBA data have a resolution of 7{\arcsec} at 667~GHz and
13{\arcsec} at 353~GHz. At these lower resolutions
contamination from dust is likely.

\section{Results}

\begin{figure}
\plotone{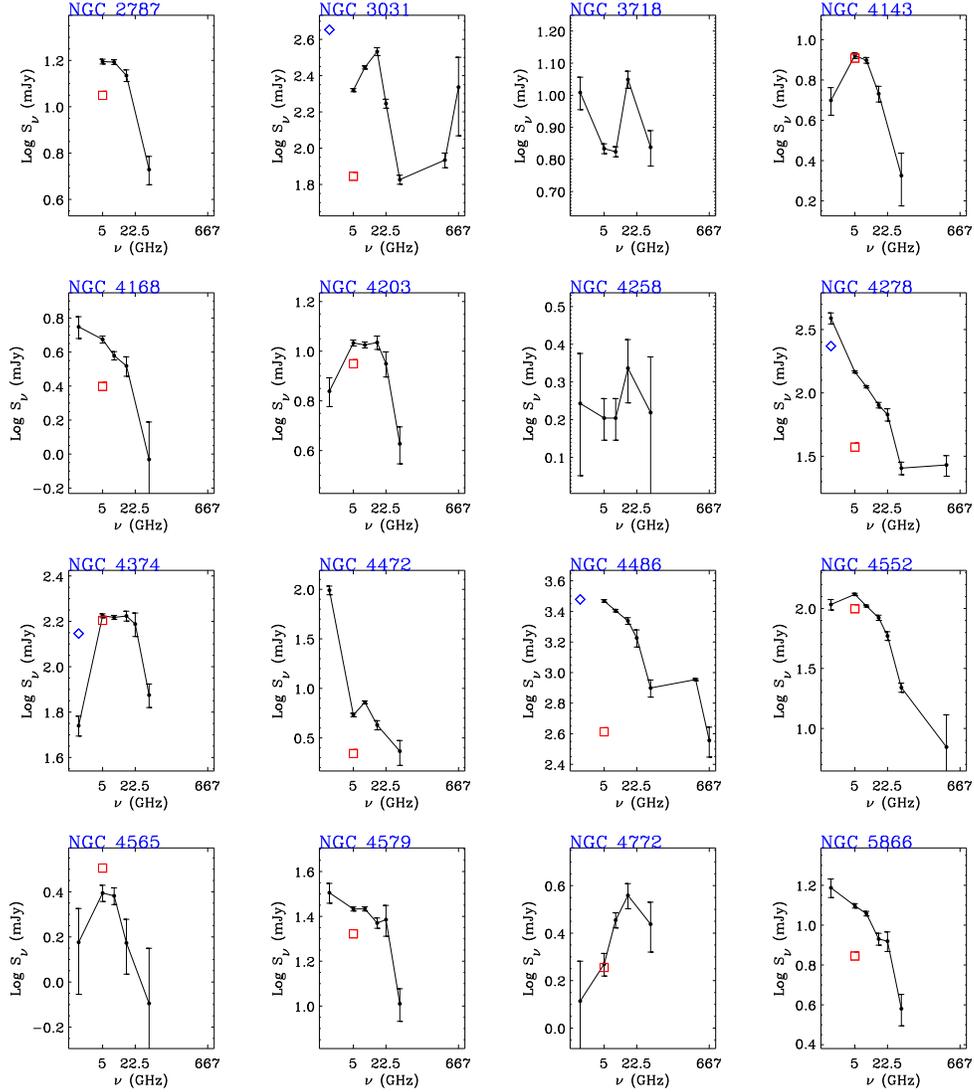}
\vspace{-1.5in}
\caption{The solid lines show the 1.4~GHz to 667~GHz radio spectral
shape of all 16 LLAGNs at D~$<$ 19~Mpc with pc-scale radio cores. 
Estimated 2$\sigma$ error bars are shown. 
The 1.4~GHz fluxes are from FIRST data. When available, MERLIN 1.6~GHz 
and VLBA 5~GHz fluxes are shown with diamonds and rectangles, respectively.}
\end{figure}

Radio spectra of the sixteen nuclei are shown in Fig.~2.
The 1.4~GHz data is from the 5{\arcsec} resolution FIRST data, 
and if a 0{\farcs}1 resolution MERLIN (preliminary) flux exists, it is 
shown by the diamond at 1.6~GHz. 
%If a 5~GHz mas-scale 
%flux exists, it is shown by the rectangle at 5~GHz. 
Given the higher resolution of MERLIN we would expect its flux measurement
to be lower than that of FIRST, unless there is significant variability.
The presence of some higher than expected MERLIN fluxes is a matter of
concern and we are looking more closely into the mapping and flux 
calibration of these data.
The arcsec-scale spectra in Fig.~2 probably closely track that of the 
mas-scale radio emission given that the peak flux-density in the 
0{\farcs}5 resolution, 6~cm VLA maps is $\sim$40\%--100\% of the total 
flux seen in the (non-simultaneous) 2.5~mas resolution, 6~cm VLBA maps 
(the rectangular symbol at 5~GHz). Also, the radio cores show similar
fluxes in 8.4~GHz and 15~GHz maps made at the highest resolution
(0{\farcs}3 and 0{\farcs}15, respectively) as in maps tapered to 
0{\farcs}5 resolution (Nagar et al. 2001). Finally, the 
high-resolution 22~GHz and 43~GHz maps show only an unresolved core with
no evidence of extended structure.

Overall the spectra show a steep fall at the high GHz frequencies.
The single exception, NGC~3031, is a nearby bright spiral with
significant nuclear dust; dust emission can explain the upturn
in its sub-mm spectrum.
Several nuclei show signs of a spectral peak in the range 1.4--20~GHz,
similar to that found for a few ellipticals by Di~Matteo et al. (2001).
This low frequency turnover is more believable since
the 1.4~GHz data have lower resolution than the 5~GHz--43~GHz data.
The spectral index above the ``peak'' is $-$0.4 to $-$1.4
(median $-$0.8), and the 
spectral index below the ``peak'' is $+$0.8 to $-$0.4
(median $\sim$0).                                         

\section{Discussion}

First, we list the main caveats:
(a)~as explained above the 5--15~GHz and the 22--43~GHz data
   have been normalized by the repeat 8.4~GHz observations.
   This assumes that the 5--43~GHz spectral shape did not change between
   the two epochs;
(b)~the 1.4~GHz, 1.6~GHz, and the 353-667~GHz data are non-simultaneous
   with the other data; 
(c)~the resolution varies between 0{\farcs}05 for the 43~GHz data and
   13{\arcsec} for the JCMT/SCUBA data. However, as discussed above,
   the radio cores are highly compact until mas-scales, so resolution 
   effects should not cause significant bias at $\geq$50~mas resolutions.

Despite these caveats, several deductions can be made from the results.
First, let us restrict ourselves to scenarios in which 
the radio core is highly compact, e.g. advection dominated 
(ADAF) or convection (CDAF) dominated accretion flows
(Narayan, Igumenshchev, \& Abramowicz 2000).
In these models, the radio emission is posited to come from the inner
10$^2$-10$^3$ Schwarzschild radii, i.e. highly sub-parsec for the
black hole masses considered here.
Thus, the radio fluxes we measure at all of our frequencies represent
upper limits to the radio emission from such models.
In fact in cases like this, it is not relevant to determine spectral
indices from matched resolution radio maps; all that accomplishes is 
a deterioration of the few high resolution datapoints.
We see no evidence of a moderately to highly inverted
radio spectrum, as predicted by most ADAF and CDAF models
(e.g. Quataert \& Narayan 1999; Di~Matteo et al. 2001; summarized
in Nagar et al. 2001). This tells us that even if such accretion
flows exist, the sub-arcsecond radio emission upto 43~GHz 
(possibly all the way to 667~GHz) is dominated by other components.

On the other hand there is significant evidence for jets dominating
the sub-arcsec radio emission in LLAGNs.
The 5 nuclei with the highest core mas-scale fluxes all show 
extended emission, reminiscent of parsec-scale jets (see Section 1).
In the three best studied nuclei - M~87, NGC~3031, and NGC~4258 - the `jet'
dominates the `core' radio emission (see Nagar et al. 2001 for  a
discussion of this). 
The radio spectral shape is also consistent with jet models:
the high frequency spectral shape is consistent
with optically-thin synchrotron emission, and the low frequency 
turnover could be caused by synchrotron self-absorption or
by free-free absorption.

It is suggestive that the LLAGNs in Fig.~2 show several properties
in common with gigahertz-peaked spectrum sources (GPSs; see O'Dea 1998 
for a review):
(a)~there is some evidence for a peak in the 1--20~GHz range;
(b)~milli-arcsec imaging often shows a 'core-jet' or symmetric parsec-scale 
    jets;
(c)~LLAGNs in Fig.~2 have sub-pc scale radio jet extents and spectral peaks 
    in the 5-10~GHz range. They thus extend to higher frequencies the 
    relationship between linear 
    size and turnover frequency in GPSs and Compact Steep Spectrum Sources 
    (CSSs; see O'Dea et al. 1998).
(d)~LLAGNs have a low X-ray luminosity (e.g. Ho et al. 2001), similar to 
    the case of GPS galaxies.        \newline
In GPS sources, the radio emission is posited to come from the expanding
lobes of the pc-scale radio jets. The low frequency turnover is deduced to
be the result of synchrotron self-absorption or free-free absorption.

%\section{Summary}
%It is now firmly established that at least some LLAGNs are genuinely
%scaled down versions of more powerful AGNs. Our multifrequency 
%investigation of sixteen of the nearest LLAGNs with confirmed
%pc-scale radio cores has shown revealed the following:
%the spectra fall off at high frequencies with spectral indices consistent
%with optically-thin synchrotron emission;
%several nuclei show a `peak' in the 1--20~GHz range;
%the low frequency turnover could be caused by synchrotron self-absorption
%or by free-free absorption.
%Overall, the morphology and spectra of radio cores in LLAGNs are
%well-explained by synchrotron emission from pc-scale jets.
%The GHz radio spectrum of LLAGNs show several similarities with
%that of more powerful Gigahertz Peaked Sources (GPSs).


\begin{references}

\reference Barth, A. J., Filippenko, A. V., \& Moran, E. C.  1999, \apj, 525, 673  

\reference Barth, A. J., Ho, L. C., Filippenko, A. V., \& Sargent, W. L. W.  1998, \apj, 496, 133  

\reference Bietenholz, M. F., Bartel, N., \& Rupen, M. P. 2000, \apj, 532, 895

\reference Braatz, J., Wilson, A. S., \& Henkel, C. 1997, \apjs, 110, 321  

\reference Di~Matteo, T., Carilli, C. L., \& Fabian, A. C.  2001, \apj, 547, 731

\reference Falcke, H. \& Biermann, P. L. 1999, \aap, 342, 49   

\reference Falcke, H., Nagar, N. M., Wilson, A. S., \& Ulvestad, J. S. 2000, \apj, 542, 197 

\reference Frank, J., King, A., \& Raine, D. 1995, in Accretion Power in Astrophysics, 2nd edition, (Cambridge: Cambridge Univ. Press)   

\reference Heckman, T. M. 1980, \aap, 87, 152  

\reference Ho, L. C., Filippenko, A. V., \& Sargent, W. L. W. 1997a, \apjs, 112, 315 (H97a)                                  

\reference Ho, L. C., Filippenko, A. V., Sargent, W. L. W., \& Peng, C. Y.  1997b, \apjs, 112, 391  

\reference Jones, D. L., Wrobel, J. M., \& Shaffer, D. B. 1984, \apj, 276, 480 

\reference Junor, W., \& Biretta, J. A. 1995, \aj, 109, 500

\reference Maoz, D., Filippenko, A. V., Ho, L. C., Rix, H.-W., Bahcall, J. N., Schneider, D. P., \& Macchetto, F. D. 1995, \apj, 440, 91

%\reference Nagar, N. M., Falcke, H., Wilson, A. S., \& Ho, L. C.  2000, \apj, 542, 186 (Paper~I)

\reference Nagar, N. M., Wilson, A. S., \& Falcke, H. 2001, \apjl, 559, 87

\reference Nagar, N. M., et al. 2002, in preparation; will appear in \aap   

% cdafs
\reference Narayan, R., Igumenshchev, I. V., \& Abramowicz, M. A. 2000, \apj, 539, 798 

%\reference Narayan, R., Mahadevan, R., \& Quataert, E. 1998, in
%The Theory of Black Hole Accretion Discs, ed.  M. A. Abramowicz,
%G. Bj\"{o}rnsson, \& J. E. Pringle (Cambridge: Cambridge Univ. Press), 148

\reference O'Dea, C.~P.\ 1998, \pasp, 110, 493 

\reference Quataert, E., \& Narayan, R. 1999, \apj, 520, 298

%\reference Sadler, E. M., Jenkins, C. R., \& Kotanyi, C. G. 1989, \mnras, 240, 591

\reference White, R. L., Becker, R. H., Helfand, D. J., \& Gregg, M. D. 1997, \apj, 475, 479

\reference Wrobel, J. M., Walker, R. C., \& Bridle, A. H.  1996, in Extragalactic radio sources: proc. of the 175th Symposium of the IAU, ed. R. D. Ekers, C. Fanti, \& L. Padrielli (Kluwer Academic Publishers), 131

\end{references}
\end{document}